\documentclass[11pt]{article}
 \pdfoutput=1
 \usepackage[margin=1.25in]{geometry}
\usepackage{amsmath}
\usepackage{times}
\usepackage{graphicx}
\usepackage{multirow}
\usepackage{libertine}
\usepackage[libertine]{newtxmath}
\usepackage[font=footnotesize,labelfont=bf]{caption}

\usepackage{graphicx}
\usepackage{subcaption}

\title{Local learning rules to attenuate forgetting in neural networks}
\author{Michael Deistler\textsuperscript{1}, Martino Sorbaro\textsuperscript{2,3}, Michael E. Rule\textsuperscript{2}, Matthias H. Hennig\textsuperscript{2,*}}

\date{\today}

\newcommand{\ve}{\mathbf}

\begin{document}

\maketitle

%
\begin{flushleft}
\bigskip
\textbf{1} Technical University of Munich, Germany \\
\textbf{2} Institute for Adaptive and Neural Computation, School of Informatics, University of Edinburgh, UK \\
\textbf{3} Computational Science and Technology, KTH Royal Institute of Technology, Stockholm, Sweden \\

\bigskip
* m.hennig@ed.ac.uk
\end{flushleft}

\section*{Abstract}

Hebbian synaptic plasticity inevitably leads to interference and forgetting when different, overlapping memory patterns are sequentially stored in the same network. Recent work on artificial neural networks shows that an information-geometric approach can be used to protect important weights to slow down forgetting. This strategy however is biologically implausible as it requires knowledge of the history of previously learned patterns. In this work, we show that a purely local weight consolidation mechanism, based on estimating energy landscape curvatures from locally available statistics, prevents pattern interference. Exploring a local calculation of energy curvature in the sparse-coding limit, we demonstrate that curvature-aware learning rules reduce forgetting in the Hopfield network. We further show that this method connects information-geometric global learning rules based on the Fisher information to local spike-dependent rules accessible to biological neural networks. We conjecture that, if combined with other learning procedures, it could provide a building-block for content-aware learning strategies that use only quantities computable in biological neural networks to attenuate pattern interference and catastrophic forgetting. Additionally, this work clarifies how global information-geometric structure in a learning problem can be exposed in local model statistics, building a deeper theoretical connection between the statistics of single units in a network, and the global structure of the collective learning space.

\section*{Significance}

How can neural networks avoid interference and forgetting when sequentially learning different yet overlapping memory patterns? In artificial neural networks, this problem has been solved using the geometric structure of parameter space conveyed by the Fisher information matrix (FIM), which reveals weights in the network that are important for encoding previously learned patterns. However, these weight consolidation rules are biologically implausible as they require global information about the parameter space and the history of learned patterns. Here we show mathematically and in simulations that an attractor network can approximate such learning rules with locally available information. This work suggests a novel interpretation of weight-dependent synaptic modifications observed experimentally, and purely local learning rules that mitigate against catastrophic forgetting in artificial neural networks.


\section*{Introduction}
Artificial Neural Networks (ANNs) have become adept at solving both supervised and unsupervised machine-learning tasks. Unlike biological neural networks however, ANNs are vulnerable to catastrophic forgetting \cite{mccloskey1989catastrophic}: ANNs forget their original trained structure if re-trained on new inputs. Recent studies have addressed catastrophic forgetting by constraining learning through globally-computed information about the importance of network parameters \cite{poole2017intelligent, liu2018rotate, serra2018overcoming, rosenfeld2017incremental, aljundi2017memory, rannen2017encoder, li2017learning, kirkpatrick2017overcoming}. However, biological neural networks must achieve the same through locally available information: neither the backpropagation algorithm \cite{bengio2015towards}, nor the creation of new units \cite{sorrells2018human}, nor non-local calculations of weight importance, can be implemented in biological networks as we currently understand them.


Here we introduce an approach that requires no information about previously stored memories and uses the measure of importance not as part of a loss function, but as a scaling factor for the learning rate. Addressing catastrophic forgetting in sequential learning in a Hopfield network, we derive a local Hebbian learning rule that calculates weight importance via a simple weight transformation. We show that this transformation is equivalent to computing Fisher Information Matrix (FIM) entries in a statistical model and that it provides a biologically plausible means to implement FIM-based solutions to catastrophic forgetting \cite{kirkpatrick2017overcoming,poole2017intelligent,liu2018rotate}.

\section*{Results}

\subsection*{Hopfield Networks}

A Hopfield network is a network consisting of $M$ binary nodes $x_i$, which are fully connected through symmetric weights $w_{ij}$. We use this network to store and retrieve a set of patterns $\ve{p}^1\dots \ve{p}^N$, with $p^n_i \in \{0, 1\}$. The sparsity of these patterns $s$ is defined as the ratio of bits being 1: $s = \langle p \rangle$. Classically \cite{Hopfield1982,tsodyks1988enhanced}, Hopfield networks are trained using a local Hebbian learning rule, in which the weights are set to
\begin{equation}
w_{ij} = \frac{1}{N} \sum_{n=1}^N (p^n_i-s)(p^n_j-s) = \frac{1}{N} \sum_{n=1}^N \xi^n_i \xi^n_j,
\label{weight_definition}
\end{equation}
with $N$ the number of stored patterns, and where we defined $\xi^n_i{=}p^n_i{-}s$. Encoding patterns in terms of $\xi$ instead of $p$ guarantees the mean of all encoded patterns is 0, as required for optimal pattern separation \cite{tsodyks1988enhanced}. The capacity of a Hopfield Network with a sparsity of $s{=}0.5$ is about $0.138{\times}M$ \cite{amit1985storing}, while sparser patterns lead to a higher capacity of the network, proportional to $(s|\ln(s)|)^{-1}$ \cite{tsodyks1988enhanced}.
Using the learning rule given above, we initialize the network by inducing local minima into the energy surface corresponding to our stored patterns. The energy of a given pattern $x$ is defined as
\begin{equation*}
E(x) = -\sum_{i,j=1}^M x_i x_j w_{ij}.
\end{equation*}
Given a weight matrix, if $x^t$ is the network state at time $t$, the network dynamics is defined as
\begin{equation}
x_i^{t+1} = \operatorname{\Theta}\left(\sum_{j=1}^M w_{ij} x^t_j - \theta\right),
\label{eqn:activity-dynamics}
\end{equation}
where $\Theta$ is the Heaviside step function and $\theta$ is a bias accounting for the sparsity, known as the neural threshold \cite{tsodyks1988enhanced}.
Repeating this several times, either synchronously for all neurons or asynchronously for a randomly chosen neuron, leads the network to converge into the energy minimum closest to its initial configuration.

\subsection*{Parameter importance and sequential learning}
In real-world learning, an agent is not presented all at once with all the information it needs to remember, nor does it have the chance of interleaving training on one memory with training on another and vice versa. Memories may have to be stored, and can be stored, one after the other, in sequence. If we want to investigate sequential learning in a Hopfield network, we can introduce an incremental rule
\begin{equation}
\Delta w_{ij} =  \eta \cdot \left( \frac{1}{N}\sum_{n=1}^N \xi^n_i \xi^n_j - w_{ij} \right)
\label{sequential_hebbian}
\end{equation}
with learning rate $\eta$. It's easy to see that following this rule, the weight matrix will converge to the value given in Equation (\ref{weight_definition}). The problem of this approach is that, while the new pattern is learned, the weight matrix is eventually entirely overwritten, and the previously stored patterns are forgotten \cite{nadal1986networks}.

The general idea we are aiming to implement to address this problem is that parameters that are particularly ``important'' for retrieving stored memories should be changed at a lower rate, or left untouched, by learning additional patterns. This will be successful if the energy landscape is highly anisotropic with respect to parameter changes, which usually is the case in ANNs as they are overdetermined. Then, some parameter changes (or combinations) cause a strong change, while others have little or no effect and can be used for new patterns. This defines sensitive and insensitive directions in parameter space, which we expect to exist in a network not saturated close to capacity, where energy minima would be quite close to each other.

\subsection*{Fisher Information in a Hopfield network}

In probabilistic models, the Fisher Information Matrix, which describes the geometry of the parameter space, can provide a measure of ``sloppiness'' for the model parameters, indicating the level of plasticity a certain weight can have. This allows learning to focus on relatively unimportant parameters, leaving important or ``stiff'' weights undisturbed. These terms, introduced in a more general context by Gutenkunst et al. \cite{gutenkunst2007universally}, define the parameter space anisotropy we want to exploit here to prevent forgetting. It is not immediately intuitive how to define the concept of Fisher Information, which applies to parameter-dependent probability distributions, in the case of deterministic Hopfield networks, where dynamics exist that drive activity into the attractor states that correspond to stored memories. However, a consistent definition can be attained by realising that the Hopfield system is equivalent to the fully visible Boltzmann Machine (FVBM) at the zero-temperature limit.

Given a probability distribution $P_{w}$ dependent on a matrix of parameters $\{w_{ij}\}$, the Fisher information matrix is defined as
\begin{equation}
F_{w_{ij},w_{kl}} = \bigg\langle \frac{\partial \log P_{w}(x)}{\partial w_{ij}} \frac{\partial  \log P_{w}(x)}{\partial w_{kl}} \bigg\rangle,
\label{Fisher_information_definition}
\end{equation}
where the brackets $\left<\cdot\right>$ denote averaging over all patterns stored in the network. Since we are interested in a measure of single-parameter importance, we consider only the diagonal of the FIM: $F_{ij} \equiv F_{w_{ij},w_{ij}}$, which then expresses the sensitivity of the distribution to changes in the weight $w_{ij}$.

The FVBM, a model identical to the fully connected Ising model, has the form
\begin{equation}
\label{eq:PME}
P_w(x) = \frac{1}{Z(T)}\exp\left( \frac{1}{T}\sum_{i \neq j}x_ix_jw_{ij} \right).
\end{equation}

The FIM can be computed from a sample extracted from $P_w$, rather than from its analytical expression, exploiting the fact that (see Appendix A):
\begin{equation}
\label{eq:fisher_cov}
F_{w_{ij},w_{kl}} = \mathrm{Cov}\big[x_i x_j, x_k x_l\big], \quad\text{ and therefore }\quad
F_{ij} = \mathrm{Var}\big[x_i x_j]
\end{equation}

In the case of a Hopfield network, however, no probability distribution of states exists, since the dynamics of the model is limited to convergence to attractors. Yet, the FVBM probability distribution (\ref{eq:PME}) converges, in the $T\to 0$ limit, to null probability for all patterns except the ones of lowest energy.
This coincides with the equilibrium distribution of a Hopfield network, which has finite probability on attractors (learned or spurious) and zero probability elsewhere. Analogously, it can be shown that the dynamics of the FVBM, for example the one defined by Monte Carlo sampling, are equivalent to the Hopfield time evolution defined in equation (\ref{eqn:activity-dynamics}).

Assuming the network parameters are set such that no spurious attractors exist, the stable patterns coincide with learned patterns in a trained network. This allows computing the variance in (\ref{eq:fisher_cov}) over this distribution, and the FIM as
\begin{equation}
\label{eq:fisher_hebbian}
F_{ij} = \frac{1}{N}\sum_{n=1}^N(\xi_i^n\xi_j^n)^2 - \frac{1}{N^2}\left(\sum_{n=1}^N \xi_i^n\xi_j^n\right)^2.
\end{equation}
The importance of each weight, computed in an appropriate way as $\Omega_{ij}=f(F_{ij})$, with $f$ being a monotonically decreasing function, can then be used to scale the learning rate in order to protect stored memories:
\begin{equation}
\label{eq:new_learning_rule}
\Delta w_{ij} = \eta \Omega_{ij} (\xi_i \xi_j - w_{ij}).
\end{equation}

\subsubsection*{A biologically plausible learning rule}

In order to evaluate the importance of a connection locally, the network has to constantly compute the sum in equation (\ref{eq:fisher_hebbian}), which requires constant sampling of previously memorized patterns. This process is not impossible: the recall and replay of memories, for example during sleep, has been both experimentally observed and theoretically studied  as a means of memory consolidation \cite{wilson1994reactivation, stickgold2005sleep}. However, we will here show that there is an even simpler local way of estimating importance from the value of the weight, at least for a Hopfield network.

We use (\ref{eq:fisher_cov}) to write the diagonal entries of the Fisher Information Matrix as
\begin{equation*}
F_{w_{ij},w_{ij}} = F_{ij} = \mathrm{Var}\big[x_i x_j\big] = \big\langle x_i^2 x_j^2 \big\rangle - \big\langle x_i x_j \big\rangle ^2.
\end{equation*}

We would like an expression for the diagonal of the FIM that depends only on locally-available weight information. We can use the fact that $w_{ij} = \left< x_i x_j \right>$ by construction (\ref{eq:fisher_cov}) to write
\begin{equation}
\begin{aligned}
F_{ij}
&= \left< x_i^2 x_j^2 \right> - w_{ij}^2,
\end{aligned}
\end{equation}
but the question remains of how to estimate $\left< x_i^2 x_j^2 \right>$, which is a fourth moment of the activity distribution. It is possible (Appendix B) to expand this term as a function of means and correlations:
\begin{equation}
\begin{aligned}
\left< x_i^2 x_j^2 \right>
&=
(1-2s)^2 \left<  x_i x_j \right>
\\&\hphantom{={}} +
s (1-s) (1-2s) \left(\left< x_{i\vphantom{j}}\right> + \left<x_j\right> \right)
\\&\hphantom{={}} +
s^2(1 - s)^2
\end{aligned}
\end{equation}
However, the expected activation rates $\left< x_{i\vphantom{j}}\right>$ and $\left<x_j\right>$ are not directly accessible during sequential learning, since previously learned patterns are not `sampled' by the network during the learning process. These mean activations are correlated with the weights, and a simple closed-form approximation does not exist. Using the fact that patterns are centered at zero-mean activation rates, we can write the approximation
\begin{equation}
F_{ij}\approx s^2(1 - s)^2 + (1-2s)^2 w_{ij} - w_{ij}^2.
\label{eqn:Hebbian_Fisher_Info}
\end{equation}
In two limiting cases, the dependence on the expected rates vanishes, and Eq. (\ref{eqn:Hebbian_Fisher_Info}) holds with equality. At $s{=}0.5$, $F_{ij}{=}\frac{1}{16} {-}w_{ij}^2$, and at $s{=}0$, $F_{ij}{=}w_{ij}(1{-}w_{ij})$. We now focus on learning near the sparse-coding limit, with $s\ll0.5$.
To simplify the online, local estimation of weight importance, we consider an approximation for a perturbation around the sparse ($s\rightarrow0$) limit.
Expanding Eq. (\ref{eqn:Hebbian_Fisher_Info}) to first-order in $s$ gives
\begin{equation}
F_{ij} \approx w_{ij} (1-w_{ij})
-4s w_{ij}
\label{sparsity_zero_fisher}
\end{equation}
%

\begin{figure}
  {\centering
	\begin{subfigure}[b]{\linewidth}
    	\includegraphics[width=\linewidth]{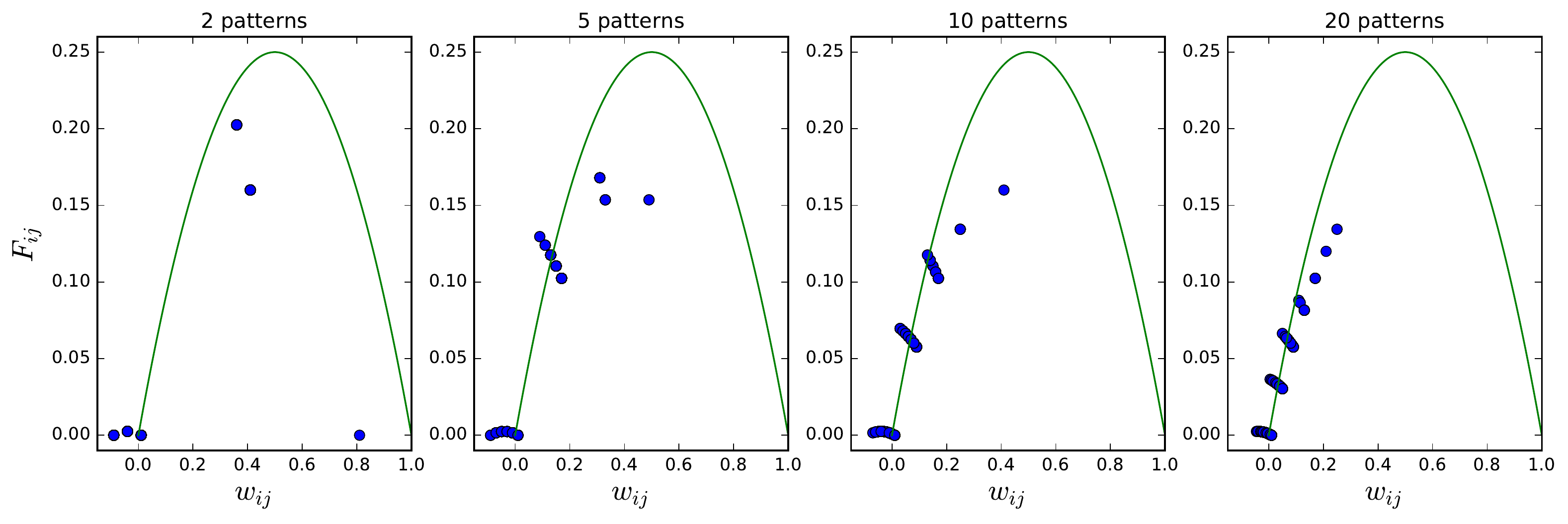}
	\end{subfigure}
    \begin{subfigure}[b]{\linewidth}
    	\includegraphics[width=\linewidth]{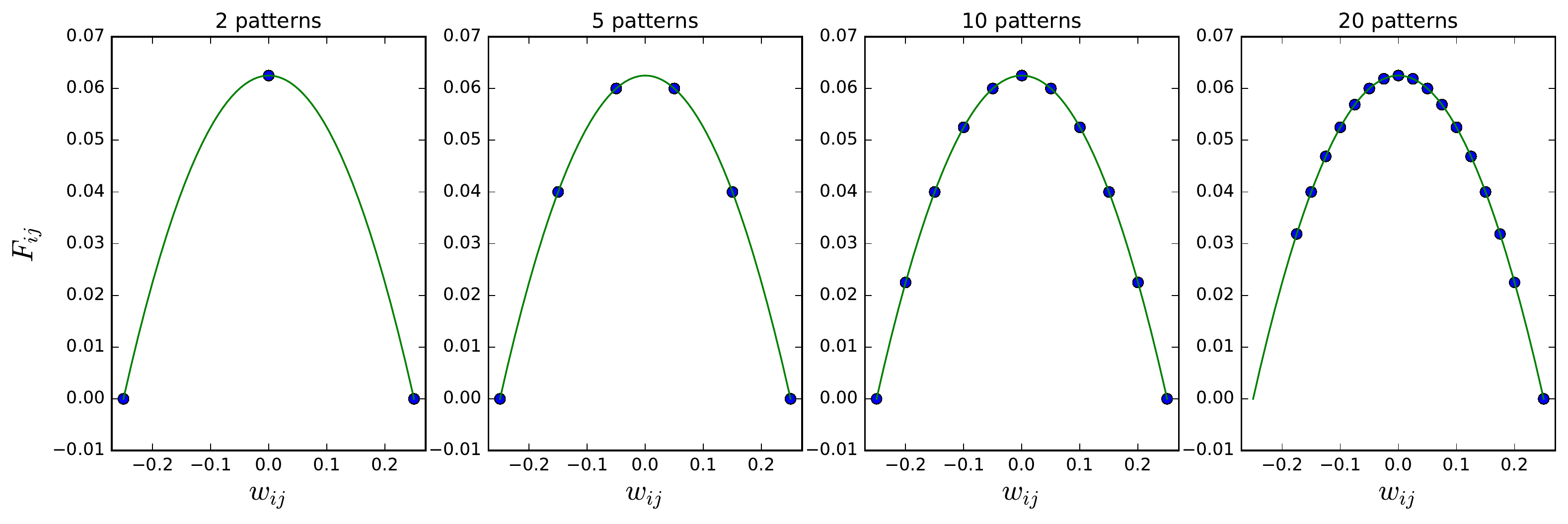}
	\end{subfigure}}
    \caption{The Fisher information-based measure of importance as a function of the weight value in a network. Blue dots refer to the true values, computed according to equation \ref{eq:fisher_cov}, green lines to the theoretically derived relation (\ref{eqn:Hebbian_Fisher_Info}). Top: measurements for $s=0.1$ compared to the analytically available relation for $s\rightarrow0$, Equation \ref{sparsity_zero_fisher}. Bottom: sparsity $s=0.5$.}
\label{weight_fisher_measure}
\end{figure}

Figure \ref{weight_fisher_measure} illustrates this approximation. Assuming that the weights are in the range $w\in[0,1]$ for $s\rightarrow0$, there is a reasonable correspondence between the predicted and actual Fisher information for $s=0.1$, and the relationship is exactly reproduced for $s=0.5$. Once several sparse patterns are stored, only values to the left of the maximum of the parabola given by Equation (\ref{sparsity_zero_fisher}) appear, an effect that becomes increasingly evident as more patterns are stored.
This is expected because, in order to have $w_{ij} = \frac{1}{N}\sum \xi_i^n\xi_j^n > 0.5$ and assuming $\zeta_i \zeta_j \in \{0,1\}$ in the limit $s \rightarrow 0$, we need more than half of the connections to be \begin{equation}
\xi_i^n\xi_j^n = 1.
\end{equation}
The probability for this to happen is
\begin{equation}
p(\zeta_i \zeta_j = 1) = p(\zeta_i = 1,  \zeta_j = 1) = p(\zeta_i = 1) p(\zeta_j = 1) = s^2.
\end{equation}
When storing $N$ patterns, at least $\frac{N}{2}$ have to be 1, which is a process that can be described by the Cumulative Distribution Function of the binomial distribution, for which no analytical solution exists. In table \ref{tab:binocdf}, some numerical values are shown.\\
As it can be seen, the probabilities for the weight being above 0.5 is decreasing with the number of patterns and is negligible small for a sufficient number of patterns.
\begin{table}
\begin{center}
    \begin{tabular}{| l | l | l | l | l |}
    \hline
    s & 2 patterns & 5 patterns & 10 patterns & 20 patterns \\ \hline
    0.05 & 6.2500e-06 & 1.5566e-07 & 5.0848e-14 & 0 \\ \hline
    0.1 &  1.0000e-04 & 9.8506e-06 & 2.0289e-10 & 0 \\
    \hline
    \end{tabular}
\end{center}
\caption{Estimated Probabilities of $w_{ij} > 0.5$ for 2, 5, 10 and 20 patterns and sparsities $s=0.05$ and $s=0.1$.}
\label{tab:binocdf}
\end{table}

These results show that for a model storing sparse patterns, the weight sensitivities increase monotonically with the value of the weight. This relationship is well captured even with a linear function. This allows constructing heuristic, fully local and hence biologically plausible learning rules that only modify irrelevant weights during continuous learning. To this end, we can generalize Equation (\ref{eq:new_learning_rule}) to introduce an additional correction to the learning rate $\Omega_{ij}$ depending on the weight value.

In the following we investigate two approaches for learning rate correction. The first consists of imposing a threshold $\Theta_w$ on each weight:
    \begin{equation}
    \Omega_{ij} =
    \begin{cases}
    1, & \text{for } w_{ij} \leq \Theta_w\\
    0, & \text{for } w_{ij} > \Theta_w
    \end{cases}
    \label{weights}
    \end{equation}
Second, the relation between Fisher information and weight in Equation \ref{sparsity_zero_fisher} suggests any strictly positive, monotonically decreasing function of the weight should provide an appropriate learning rate correction. An interpretation of the weight value as the curvature of a Gaussian approximation to the weight posterior predicts an inverse relationship (see Appendix C for derivation). In simulations we however found a better performance using an exponential scaling of the weight with
    \begin{equation}
    \Omega_{ij} = \exp(-a |w_{ij}|).
    \label{decreasing}
    \end{equation}
Following the Bayesian interpretation, this rule can be further augmented by only updating weights with strong changes:
\begin{equation}
\Omega_{ij} =
    \begin{cases}
    \exp(-a |w_{ij}|) & \text{for } \Delta w_{ij}>\Theta_{\Delta w}\\
    \Omega_{ij} = 0 & \text{else}
    \end{cases}
    \label{decreasing-high}
\end{equation}
In addition to not modifying important weights, this will prevent weight changes that do not support the new pattern. In the following we show in simulations that these rules indeed prevent overwriting of stored memories, and enable continuous learning in the Hopfield network.

\subsection*{Simulations}

\begin{figure}
{\centering
  \includegraphics{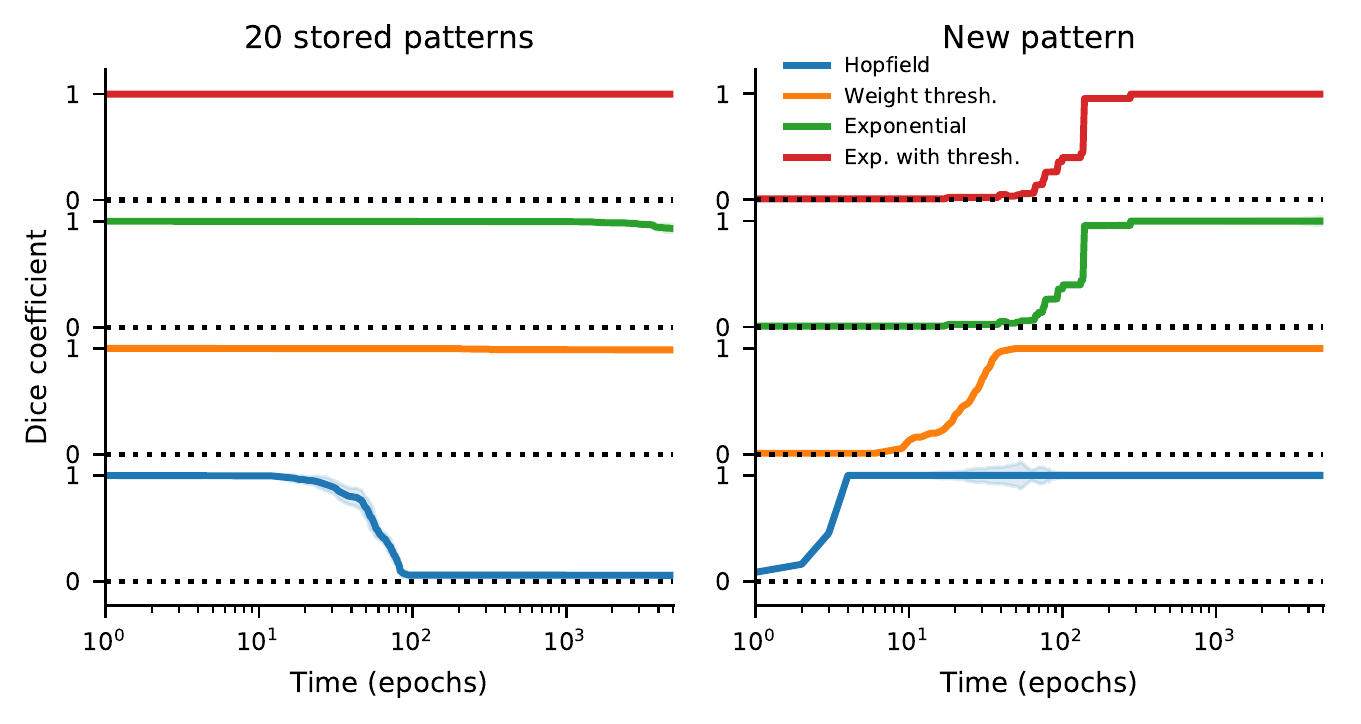}
  \caption{Modified local Hebbian learning rules prevent catastrophic forgetting. A network of 100 neurons was initialised with 20 patterns, and a novel pattern was learned with the incremental rule (Equation \ref{sequential_hebbian}), and augmented versions of this rule. The learning rate was $\eta=0.01$, curves show the average Dice coefficient from 20 simulations. Parameters for the augmented learning rules: weight threshold, Eqn. \ref{weights} - $\Theta_w=0.001$; exponential, Eqn. \ref{decreasing} - $a=220$; exponential with threshold, Eqn. \ref{decreasing-high} - $a=220$ and $\Theta_{\Delta w}=0.2*\eta$}
  \label{fig:one_new_pattern}
  }
\end{figure}

We first consider a network with 20 patterns stored using Equation \ref{weight_definition}, and a new pattern learned with the incremental rule (Eqn. \ref{sequential_hebbian}). Pattern retention is assessed by the S{\o}rensen-Dice coefficient $D = 2 \cdot |A \cap B|/(|A|+|B|)$, where $A$ and $B$ reflect the binary bit-vectors reflecting a target (true) pattern and a recovered (stored) pattern, computed at each learning rule iteration by synchronously applying Equation \ref{eqn:activity-dynamics} ten times.

As expected, the incremental rule rapidly removes all traces of the previously stored patterns, while the new pattern is reliably stored. Reducing the learning rate $\eta$ can increase retention as overwriting is slower, but this also slows down the learning of the new pattern, always resulting in exponentially fast forgetting \cite{barrett2008optimal}. In contrast, for the augmented local rules the stored patterns are retained. Since the modified rule effectively shows the learning rate, storing the new pattern is slower than during normal Hopfield learning. Importantly, in these simulations the learning rate correction is re-computed at each iteration based on the current weights. Therefore successful retention is possible even for a rule operating entirely locally on each weight and in time.

\begin{figure}
{\centering
  \includegraphics{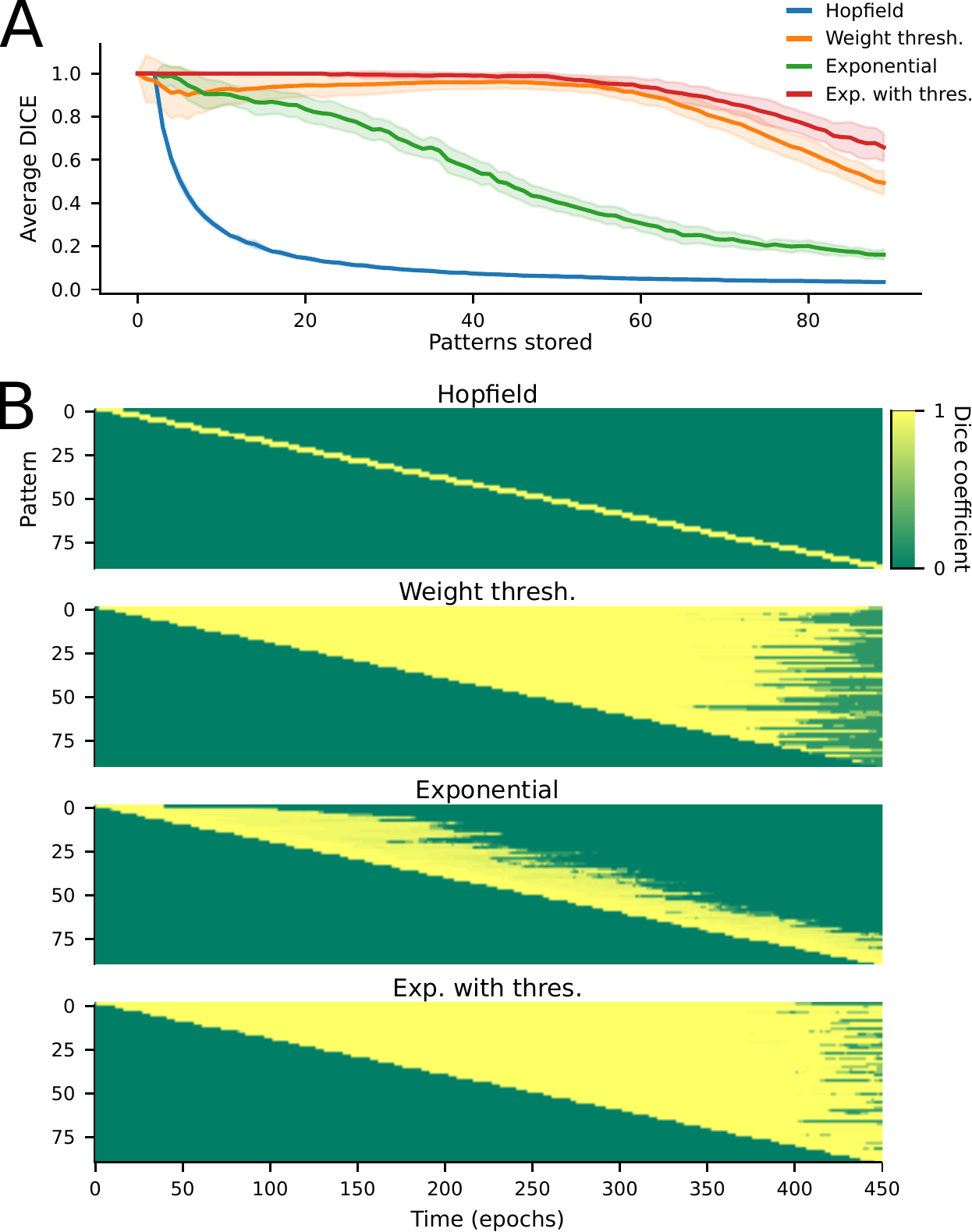}
  \caption{Continuous learning without catastrophic forgetting. In a 100 unit network, pre-trained with 20 patterns (disregarded for evaluation and not shown in these figures), 80 novel patterns were learned in succession. Each pattern was presented for five epochs, with learning rate $\eta=0.1$, and with parameters optimized to maximize recall for 60 patterns. {\bf A} The average DICE coefficient for all previously stored patterns, computed every time a new pattern was stored, using different learning rules. The averages are for 20 simulations, and shaded areas indicate the standard deviation. {\bf B} The average Dice coefficient for each pattern as a function of the training epoch, for different learning rules. The simulations show that augmented learning rules have improved retention, compared to the normal Hebb rule, but their behavior differs with increasing network loading.}
  \label{fig:many_new_patterns}
  }
\end{figure}

Next we extend our approach to continuous learning, presenting new patterns one by one for a fixed number of iterations, and updating weights with the incremental rule. Since the Fisher information is flat for a randomly connected network, which would prevent learning with the augmented rules, the network was initialized with 20 patterns that are not tracked. For each learning rule, parameters were numerically optimized to maximize the DICE coefficient for 60 patterns with the Nelder-Mead simplex algorithm (fmin() in SciPy).

As for a single pattern, the augmented learning rules prevent forgetting also for continuous learning multiple patterns (Fig. \ref{fig:many_new_patterns}). The hard weight threshold allows loading the network up to the theoretical capacity of about 60 patterns (100 units, sparseness $s=0.1$). Beyond this capacity, all stored patterns are erased simultaneously. In contrast, the exponential rule, which continuously modifies all weights, exhibits gradual forgetting, but can, at any one time, retain a fixed number of patterns. While this behavior effectively reduces the network capacity, it also prevents catastrophic forgetting due to network overloading. Finally, when only larger weight changes beyond a fixed threshold are allowed for the exponential rule, full network loading is again possible.

\section*{Discussion}

How neurons in the brain coordinate globally to store and retrieve information remains a major open question. In particular, we do not understand how global optimization problems can be solved reliably and robustly using only local learning rules. In this work, we have explored in a Hopfield network one aspect of this global coordination: how patterns might be routed and stored in an associative memory to reduce pattern interference and catastrophic forgetting.

FIM-based approaches to continuous learning have been adopted before \cite{zenke2017continual,kirkpatrick2017overcoming,aljundi2017memory}. However, these approaches employ a regularized loss function to account for previously learned tasks. Kirpatrick et al. \cite{kirkpatrick2017overcoming} approximate the negative log-likelihood curvature using the diagonal of the Fisher Inforation Matrix (FIM) for each learned task, and Zenke et al. \cite{zenke2017continual} propose a similar approach that can be calculated online. Recently, an approach which can be generalized to be local in supervised feed-forward networks has been described by Aljundi et al. \cite{aljundi2017memory}. This approach can be generalized as a local learning rule only given certain assumptions such as a Rectified Linear Unit (ReLU) activation function. However, all of those approaches rely on implementing a regularized loss function. In contrast, scaling the learning rates alters the timescales of forgetting, but does not change the asymptotic behavior of the network: at sufficiently long timescales, the accumulated information of new patterns will still overwrite previously stored patterns. Therefore, our approach attenuates forgetting and interference in the learning phase, and might be combined with other strategies to achieve more permanent stability. In addition to immediate impact in how we understand learning in biological neural networks, local learning rules have potential to accelerate machine learning as global connectivity requirements can suffer from memory transfer bottlenecks in large-scale parallel implementations running on graphics processes and clusters.

We demonstrate our approach in a Hopfield network, a fully connected network that stores patterns via Hebbian learning, and retrieves those patterns through dynamics that minimize an energy function, moving activity into local basins of attraction \cite{Hopfield1982, cheng1994neural}. Its inherent instabilities and unlearning of previously learned patterns have attracted considerable interest in the past \cite{hopfield1983unlearning, robins1998catastrophic, french1999catastrophic}. The Hebbian learning rule and neural-like dynamics make the Hopfield network a relevant model of biological memory, although the required symmetric weights are biologically implausible. Our model however makes two specific predictions for networks that implement similar attractor dynamics, which are both supported by experimental data. First, individual synapse stability is expected to be proportional to its strength, since strong synapses are the most important retaining memories. This effect has been reported in chronic {\em in vivo} experiments monitoring cortical spines \cite{Holtmaat2005,Loewenstein2015a}. Second, as cortical networks mature, and more synapses are involved in maintaining stored memories, the proportion of stable synapses should increase. Chronic imaging during cortical development, which demonstrated an increase in the fraction of stable synapses after the critical period, confirm this prediction \cite{Grutzendler2002,Holtmaat2005}. It is interesting to note that there appears to be less synapse stability in the hippocampus \cite{attardo2015impermanence}, which may be consistent with its function as a temporary episodic memory system, and suggests potential differences in the synaptic plasticity rule.

Further evidence for protection from catastrophic forgetting in cortical networks comes from studies investigating the stability of neural activity over longer time intervals. These found a small fraction of very stable neurons, which had high firing rates and were important for stabilizing the whole network dynamics, while the remaining neurons showed considerable changes \cite{panas2015sloppiness,Grosmark2016a}. This pattern does not emerge in the Hopfield model, where the activity of neurons is more homogeneous than in cortical circuits, and suggests an additional organizing principle in cortical networks that conveys stability of acquired knowledge \cite{rogerson2014synaptic,fauth2015formation}.

In addition to biological learning, our results can be generalized to stochastic ANNs. The Hopfield model exactly replicates the behavior of a fully visible Boltzmann machine (FVBM) at zero temperature, where the structure of the weights between neurons allows only certain activity configurations corresponding to local minima in the energy landscape. Hence, the Hopfield energy function can also be interpreted as negative log-likelihood of a FVMB. In this case, a Fisher Information based learning rule will protect low-energy network configurations which correspond to high probability states. Since the learning rule protects the joint configuration of the whole network, relevant learned configurations, for instance trained through backpropagation in a deep network, are stable during continuous learning of new tasks, as demonstrated using a penalty in a global loss function by Kirpatrick et al. \cite{kirkpatrick2017overcoming}.

\section*{Acknowledgments}

Funding was provided by the Engineering and Physical Sciences Research Council grant EP/L027208/1. M.S. was supported by the EuroSPIN Erasmus Mundus Program, the EPSRC Doctoral Training Centre in Neuroinformatics (EP/F500385/1 and BB/F529254/1), and a Google Doctoral Fellowship. We thank Mark van Rossum and Jo\~ao Sacramento for comments.

\clearpage

\section*{Appendix A: Derivation of Fisher information as covariance}
\label{app:fisher_covariance}
Under certain regularity assumptions, we can rewrite the Fisher Information as
\begin{equation*}
F_{w_{ij},w_{kl}} = -\bigg\langle \frac{\partial}{\partial w_{ij}\partial w_{kl}} \log\big(p(\underline{x})\big)\bigg\rangle.
\end{equation*}
This form provides an intuitive interpretation of $F_{ij}$ as the curvature of the energy landscape. A high Fisher Information hence corresponds to a high curvature - and hence a strong change in energy when perturbing the given parameter.
in the above equation, the probability of a pattern is given by
\begin{equation}
p(\underline{x}) =  \frac{1}{Z}\exp{\big(-E\big)} = \frac{1}{Z}\exp{\Big(\sum_{n,m}x_n x_m w_{nm}\Big)}
\label{probability}
\end{equation}
with $Z$ being
\begin{equation*}
Z = \sum_{\{\underline{x}\}}\exp{\Big(\sum_{n,m}x_n x_m w_{nm}\Big)}
\end{equation*}
We can hence write the log-probability as
\begin{equation*}
\log\big(p(\underline{x})\big) = \Big(\sum_{n,m}x_n x_m w_{nm}\Big) - \log(Z)
\end{equation*}
Plugging this into (\ref{Fisher_information_definition}) leads to
\begin{equation*}
F_{w_{ij},w_{kl}} = \bigg\langle \bigg(\frac{\partial}{\partial w_{ij}} \Big(\sum_{n,m}x_n x_m w_{nm}\Big) - \log(Z)\bigg) \bigg(\frac{\partial}{\partial w_{kl}} \Big(\sum_{n,m}x_n x_m w_{nm}\Big) - \log(Z)\bigg)\bigg\rangle
\end{equation*}
We differentiate this using the chain rule
\begin{equation*}
F_{w_{ij},w_{kl}} = \bigg\langle \bigg(x_i x_j - \frac{1}{Z} \frac{\partial}{\partial w_{ij}} Z\bigg) \bigg(x_k x_l - \frac{1}{Z} \frac{\partial}{\partial w_{kl}} Z\bigg)\bigg\rangle
\end{equation*}
We differentiate $Z$
\begin{equation*}
\frac{\partial}{\partial w_{ij}} Z = \sum_{\{\underline{x}\}}\Big(\exp{\Big(\sum_{n,m}x_n x_m w_{nm}\Big)}x_i x_j\Big)
\end{equation*}
and use (\ref{probability}) leading to
\begin{equation*}
F_{w_{ij},w_{kl}} = \Big\langle \big(x_i x_j - \sum_{\{\underline{x}\}}p(\underline{x}) x_i x_j\big)\big(x_k x_l - \sum_{\{\underline{x}\}}p(\underline{x}) x_k x_l\big)\Big\rangle
\end{equation*}
which is the definition of covariance:
\begin{equation*}
F_{w_{ij},w_{kl}} = \mathrm{Cov}\big[x_i x_j, x_k x_l\big].
\end{equation*}

\clearpage

\section*{Appendix B: Expansion of FIM diagonal in terms of weights}

In order to estimate the FIM diagonal entries for the weights, we must estimate fourth moments of the activity, $\left< x_i^2 x_j^2 \right>$. The Hopfield network represents the zero-temperature limit of a pairwise spin model, which is determined entirely by the first two moments $\left<x\right>$ and $\left<x x^\top \right>$. It is therefore possible to derive an expression for this fourth moment, $\left< x_i^2 x_j^2 \right>$, in terms of means and correlations. We first expand $\left< x_i^2 x_j^2 \right>$ based on $x_i{=}p_i{-}s$ and $x_j{=}p_j{-}s$, where $p$ reflects the binary patterns being encoded, and $s$ is the sparsity level of our encoding:
$$
\left< x_i^2 x_j^2 \right>
=
\left< (p_i -s)^2 (p_j -s)^2 \right>
$$
Because $p$ is binary, $p^2=p$, and the quadratic terms simplify on expansion:
\begin{equation}
\begin{aligned}
(p -s)^2
&=
p^2 - 2 s p + s^2
\\&=
p - 2 s p + s^2
\\&=
p(1-2s)+s^2
\end{aligned}
\end{equation}
One can therefore expand $\left< x_i^2 x_j^2 \right>$ as
\begin{equation}
\begin{aligned}
\left< x_i^2 x_j^2 \right>
&=
\left< (p_i(1-2s)+s^2) (p_j(1-2s)+s^2) \right>
\\&=
(1-2s)^2 \left< p_i p_j \right> +
s^2 (1-2s) \left[ \left<p_{i\vphantom{j}}\right> + \left<p_j\right> \right] + s^4
\end{aligned}
\end{equation}
This simplification, arising the binary nature of the spins $p$, will allow us to express $\left< x_i^2 x_j^2 \right>$ in terms of lower-order moments.
We would like this expansion in terms of weights $w_{ij} = \left< x_i x_j \right>$, and so substitute $p_i{=}x_i{+}s$ and $p_j{=}x_j{+}s$
\begin{equation}
\begin{aligned}
\left< x_i^2 x_j^2 \right>
&=
(1-2s)^2 \left<  (x_i + s) ( x_j + s) \right>
\\&\hphantom{={}} +
s^2 (1-2s) \left[ \left< x_{i\vphantom{j}} + s\right> + \left<x_j + s\right> \right]
+ s^4
\\&=
(1-2s)^2 \left[
\left<  x_i x_j \right>
+ s \left(\left< x_{i\vphantom{j}}\right> + \left<x_j\right> \right)
+s^2
\right]
\\&\hphantom{={}} +
s^2 (1-2s) \left[ \left< x_{i\vphantom{j}}\right> + \left<x_j\right> + 2s \right]
+ s^4
\\&=
(1-2s)^2 \left<  x_i x_j \right>
\\&\hphantom{={}} +
s (1-s) (1-2s) \left(\left< x_{i\vphantom{j}}\right> + \left<x_j\right> \right)
\\&\hphantom{={}} +
s^2(1 - s)^2
\end{aligned}
\end{equation}
This expresses $\left< x_i^2 x_j^2 \right>$ in terms of fist moments, $\left< x_{i\vphantom{j}}\right>$, and second moments $\left<  x_i x_j \right>$. The second moments are identified with the weights $w_{ij}$ by construction (Eq. \ref{weight_definition}). The expected activations $\left< x_{i\vphantom{j}}\right>$ could be estimated via sampling, if stored patterns are re-activated, or may be approximated by their expected-value of zero.

\clearpage

\section*{Appendix C: Curvature-aware Hebbian learning}

We begin with the incremental learning rule, dropping indices for legibility,
\begin{equation}
\Delta w = \Omega ( \hat w - w ),
\label{eq:update}
\end{equation}
where $w$ is a weight, $\hat w$ is the new weight indicated by data, and $\Omega$ is a function that adjusts the learning rate. Time constants, step size, and learning rates have been absorbed into $\Omega$ in this case.
This equation can also be written by interpreting $\Omega$ as a convex combination of the old and the new weights as
\begin{equation}
w_{new} = w + \Delta w = w + \Omega ( \hat w - w ) = \Omega \hat w + (1-\Omega) w.
\end{equation}

We now consider a Bayesian update of a Gaussian approximation to the posterior state for the value of a weight $w$.
Let our current estimate have mean $w$ and precision $\tau$. Let our update have estimated weight $\hat w$ and a constant precision $c$.
We then interpret the Fisher information as the curvature (precision) of the prior, thus equating our FIM estimate with the precision $\tau$.

The Bayesian update to the mean of a Gaussian is the weighted sum
\begin{equation}
w_{new} = \frac { c \hat w  + \tau w} {c +\tau},
\frac { c  } {c +\tau} \hat w
+
\frac { \tau } {c +\tau} w.
\end{equation}
Introducing
\begin{equation}
\beta = \frac { c  } {c +\tau} = \frac 1 {1+\tfrac 1 c \tau},
\end{equation}
we can write the weight update as the convex combination
\begin{equation}
w_{new} =
\beta \hat w + (1-\beta) w.
\end{equation}
Interpreting $\tau$ as the FIM for weight $w$, and using $\tau \approx w (1-w)$, we obtain
\begin{equation}
\beta = \frac 1 {1+ \tfrac 1 c w - \tfrac 1 c w^2}.
\end{equation}



\newpage


\begin{thebibliography}{10}

\bibitem{aljundi2017memory}
R.~Aljundi, F.~Babiloni, M.~Elhoseiny, M.~Rohrbach, and T.~Tuytelaars.
\newblock Memory aware synapses: Learning what (not) to forget.
\newblock {\em arXiv preprint arXiv:1711.09601}, 2017.

\bibitem{amit1985storing}
D.~J. Amit, H.~Gutfreund, and H.~Sompolinsky.
\newblock Storing infinite numbers of patterns in a spin-glass model of neural
  networks.
\newblock {\em Physical Review Letters}, 55(14):1530, 1985.

\bibitem{attardo2015impermanence}
A.~Attardo, J.~E. Fitzgerald, and M.~J. Schnitzer.
\newblock Impermanence of dendritic spines in live adult ca1 hippocampus.
\newblock {\em Nature}, 523(7562):592, 2015.

\bibitem{barrett2008optimal}
A.~B. Barrett and M.~C. van Rossum.
\newblock Optimal learning rules for discrete synapses.
\newblock {\em PLoS Computational Biology}, 4(11):e1000230, 2008.

\bibitem{bengio2015towards}
Y.~Bengio, D.-H. Lee, J.~Bornschein, T.~Mesnard, and Z.~Lin.
\newblock Towards biologically plausible deep learning.
\newblock {\em arXiv preprint arXiv:1502.04156}, 2015.

\bibitem{cheng1994neural}
B.~Cheng and D.~M. Titterington.
\newblock Neural networks: A review from a statistical perspective.
\newblock {\em Statistical Science}, pages 2--30, 1994.

\bibitem{fauth2015formation}
M.~Fauth, F.~W{\"o}rg{\"o}tter, and C.~Tetzlaff.
\newblock Formation and maintenance of robust long-term information storage in
  the presence of synaptic turnover.
\newblock {\em PLoS Computational Biology}, 11(12):e1004684, 2015.

\bibitem{french1999catastrophic}
R.~M. French.
\newblock Catastrophic forgetting in connectionist networks.
\newblock {\em Trends in Cognitive Sciences}, 3(4):128--135, 1999.

\bibitem{Grosmark2016a}
A.~D. Grosmark and G.~Buzsaki.
\newblock {Diversity in neural firing dynamics supports both rigid and learned
  hippocampal sequences}.
\newblock {\em Science}, 351(6280):1440--1443, 2016.

\bibitem{Grutzendler2002}
J.~Grutzendler, N.~Kasthuri, and W.-b. Gan.
\newblock {Long-term dendritic spine stability in the adult cortex.}
\newblock {\em Nature}, 420(6917):812--6, 2002.

\bibitem{gutenkunst2007universally}
R.~N. Gutenkunst, J.~J. Waterfall, F.~P. Casey, K.~S. Brown, C.~R. Myers, and
  J.~P. Sethna.
\newblock Universally sloppy parameter sensitivities in systems biology models.
\newblock {\em PLoS Computational Biology}, 3(10):e189, 2007.

\bibitem{Holtmaat2005}
A.~J. G.~D. Holtmaat, J.~T. Trachtenberg, L.~Wilbrecht, G.~M. Shepherd,
  X.~Zhang, G.~W. Knott, and K.~Svoboda.
\newblock {Transient and persistent dendritic spines in the neocortex in vivo.}
\newblock {\em Neuron}, 45(2):279--91, jan 2005.

\bibitem{Hopfield1982}
J.~J. Hopfield.
\newblock {Neural networks and physical systems with emergent collective
  computational abilities.}
\newblock {\em Proceedings of the National Academy of Sciences},
  79(8):2554--2558, apr 1982.

\bibitem{hopfield1983unlearning}
J.~J. Hopfield, D.~Feinstein, and R.~Palmer.
\newblock ‘unlearning’has a stabilizing effect in collective memories.
\newblock {\em Nature}, 304(5922):158, 1983.

\bibitem{kirkpatrick2017overcoming}
J.~Kirkpatrick, R.~Pascanu, N.~Rabinowitz, J.~Veness, G.~Desjardins, A.~A.
  Rusu, K.~Milan, J.~Quan, T.~Ramalho, A.~Grabska-Barwinska, et~al.
\newblock Overcoming catastrophic forgetting in neural networks.
\newblock {\em Proceedings of the National Academy of Sciences},
  114(13):3521--3526, 2017.

\bibitem{li2017learning}
Z.~Li and D.~Hoiem.
\newblock Learning without forgetting.
\newblock {\em IEEE Transactions on Pattern Analysis and Machine Intelligence},
  2017.

\bibitem{liu2018rotate}
X.~Liu, M.~Masana, L.~Herranz, J.~Van~de Weijer, A.~M. Lopez, and A.~D.
  Bagdanov.
\newblock Rotate your networks: Better weight consolidation and less
  catastrophic forgetting.
\newblock {\em arXiv preprint arXiv:1802.02950}, 2018.

\bibitem{Loewenstein2015a}
Y.~Loewenstein, U.~Yanover, and S.~Rumpel.
\newblock {Predicting the Dynamics of Network Connectivity in the Neocortex.}
\newblock {\em Journal of Neuroscience}, 35(36):12535--12544, 2015.

\bibitem{mccloskey1989catastrophic}
M.~McCloskey and N.~J. Cohen.
\newblock Catastrophic interference in connectionist networks: The sequential
  learning problem.
\newblock In {\em Psychology of Learning and Motivation}, volume~24, pages
  109--165. Elsevier, 1989.

\bibitem{nadal1986networks}
J.~Nadal, G.~Toulouse, J.~Changeux, and S.~Dehaene.
\newblock Networks of formal neurons and memory palimpsests.
\newblock {\em Europhysics Letters}, 1(10):535, 1986.

\bibitem{panas2015sloppiness}
D.~Panas, H.~Amin, A.~Maccione, O.~Muthmann, M.~van Rossum, L.~Berdondini, and
  M.~H. Hennig.
\newblock Sloppiness in spontaneously active neuronal networks.
\newblock {\em Journal of Neuroscience}, 35(22):8480--8492, 2015.

\bibitem{poole2017intelligent}
B.~Poole, F.~Zenke, and S.~Ganguli.
\newblock Intelligent synapses for multi-task and transfer learning.
\newblock In {\em International Conference on Learning Representations}, 2017.

\bibitem{rannen2017encoder}
A.~Rannen Ep~Triki, R.~Aljundi, M.~Blaschko, and T.~Tuytelaars.
\newblock Encoder based lifelong learning.
\newblock In {\em Proceedings ICCV 2017}, pages 1320--1328, 2017.

\bibitem{robins1998catastrophic}
A.~Robins and S.~McCALLUM.
\newblock Catastrophic forgetting and the pseudorehearsal solution in
  hopfield-type networks.
\newblock {\em Connection Science}, 10(2):121--135, 1998.

\bibitem{rogerson2014synaptic}
T.~Rogerson, D.~J. Cai, A.~Frank, Y.~Sano, J.~Shobe, M.~F. Lopez-Aranda, and
  A.~J. Silva.
\newblock Synaptic tagging during memory allocation.
\newblock {\em Nature Reviews Neuroscience}, 15(3):157, 2014.

\bibitem{rosenfeld2017incremental}
A.~Rosenfeld and J.~K. Tsotsos.
\newblock Incremental learning through deep adaptation.
\newblock {\em arXiv preprint arXiv:1705.04228}, 2017.

\bibitem{serra2018overcoming}
J.~Serr{\`a}, D.~Sur{\'\i}s, M.~Miron, and A.~Karatzoglou.
\newblock Overcoming catastrophic forgetting with hard attention to the task.
\newblock {\em arXiv preprint arXiv:1801.01423}, 2018.

\bibitem{sorrells2018human}
S.~F. Sorrells, M.~F. Paredes, A.~Cebrian-Silla, K.~Sandoval, D.~Qi, K.~W.
  Kelley, D.~James, S.~Mayer, J.~Chang, K.~I. Auguste, et~al.
\newblock Human hippocampal neurogenesis drops sharply in children to
  undetectable levels in adults.
\newblock {\em Nature}, 555(7696):377, 2018.

\bibitem{stickgold2005sleep}
R.~Stickgold.
\newblock Sleep-dependent memory consolidation.
\newblock {\em Nature}, 437(7063):1272, 2005.

\bibitem{tsodyks1988enhanced}
M.~Tsodyks and M.~Feigel'Man.
\newblock The enhanced storage capacity in neural networks with low activity
  level.
\newblock {\em EPL (Europhysics Letters)}, 6(2):101, 1988.

\bibitem{wilson1994reactivation}
M.~A. Wilson and B.~L. McNaughton.
\newblock Reactivation of hippocampal ensemble memories during sleep.
\newblock {\em Science}, 265(5172):676--679, 1994.

\bibitem{zenke2017continual}
F.~Zenke, B.~Poole, and S.~Ganguli.
\newblock Continual learning through synaptic intelligence.
\newblock In {\em International Conference on Machine Learning}, pages
  3987--3995, 2017.

\end{thebibliography}
\bibliographystyle{abbrv}

\end{document}